A WIRELESS EMBEDDED TONGUE TACTILE BIOFEEDBACK SYSTEM FOR BALANCE CONTROL

Nicolas VUILLERME[1], Nicolas PINSAULT, Olivier CHENU, Anthony FLEURY, Yohan PAYAN and Jacques DEMONGEOT[1]

Laboratoire TIMC-IMAG, UMR UJF CNRS 5525, La Tronche, France

**Address for correspondence:**

Nicolas VUILLERME / Jacques DEMONGEOT

Laboratoire TIMC-IMAG, UMR UJF CNRS 5525

Faculté de Médecine

38706 La Tronche cédex

France.

Tel: (33) (0) 4 56 52 01 08

Fax: (33) (0) 4 76 76 88 44

Email: nicolas.vuillerme@imag.fr / jacques.demongeot@imag.fr

**ABSTRACT**

We describe the architecture of an original biofeedback system for balance improvement for fall prevention and present results of a feasibility study. The underlying principle of this biofeedback consists of providing supplementary information related to foot sole pressure distribution through a wireless embedded tongue-placed tactile output device. Twelve young healthy adults voluntarily participated in this experiment. They were asked to stand as immobile as possible with their eyes closed in two conditions of nobiofeedback and biofeedback. Centre of foot pressure (CoP) displacements were recorded using a force platform. Results showed reduced CoP displacements in the biofeedback relative to the no-biofeedback condition. On the whole, the present findings evidence the effectiveness of this system in improving postural control on young healthy adults. Further investigations are needed to strengthen the potential clinical value of this device.

.

# 1. Introduction

## 1.1. Balance control in older adults

Falls in older adults constitutes a major health care problem (e.g. [4,34]). More than 30% of community-dwelling persons aged 65 or more [9] and 50% of those over the age of 80 [16] fall annually and many fall more than once. In addition to the high medical expenses that falls pose to the public health service, the consequences for older adults are rather dramatic. Falls are associated with physical and psychological trauma, reduced activity, loss of independence, decreased quality of life and even injury-related deaths. For instance, the mortality of elderly nursing home residents, taken over a one-year period after falling, has been reported to be more than twice that of a non-faller control group [12]. That is the reason why prevention of falls has been an important area of research into the health of older adults.

Although falling represents a complex and multifactorial problem [17,18], the degradation of balance capacities, associated with aging (see [10] for a review), is usually considered as a major contributing factor [3,7,8,27]. Postural control requires the integration of sensory inputs to assess the position and motion of the body in space and the ability to generate forces to control body position [20]. Among the sensory inputs relevant to balance control, the importance of cutaneous information from the foot sole is well recognised (e.g., [14,23]). Indeed, plantar cutaneous mechanoreceptors (deep and plantar-surface) could potentially provide detailed spatial and temporal information about contact pressures under the foot and shear forces resulting from body movement that constitute valuable feedback to the postural control system. For instance, anaesthetising [23], altering [31] or stimulating [2,19,21,29,30,37] plantar cutaneous receptors of the plantar soles have previously been shown to affect postural control during quiet standing. In addition, as the neuromuscular constraints acting on the individual increase, as is the case following muscular fatiguing exercise, the availability and integrity of cutaneous inputs from the plantar soles become of greater importance in the appropriate control of balance [44,47]. What is more, one of the more pervasive effects of aging is the decline in plantar-surface sensitivity at various locations across the sole of the foot (e.g., [15,24,49]). The increased sensory thresholds observed in older adults were hypothesized to stem from changes in receptor morphology, reduction of receptor density, decreased elasticity of the skin and decreased nerve conduction [15]. Interestingly, altered plantar cutaneous sensation has even been identified as an important contributing factor to the occurrence of falls in the elderly [17,22]. Within this context, proposing a therapeutic intervention [37] and/or designing a technical assistance [19,29,30] to increase the somatosensory function of the plantar sole could be of great interest for improving balance and preventing falls in older adults.

## 1.2. Biofeedback systems for balance control

Biofeedback systems for balance control are designed to provide sensory information to the user when one of the sensory inputs becomes unavailable/undermined/altered, or when one merely wants to enhance his/her sensory acuity for accurate performance in daily-living, professional or sportive activities. Augmented/substituted sensory biofeedback, widely used in physical therapy and rehabilitation, is mostly delivered though visual (e.g., [28,33,50]) or acoustic sensory channels (e.g., [5,11,25]). At this point, however, these biofeedback systems, interfering ipso facto with the use of vision and hearing and hence presumably leading to a multi-tasking deficit, seem not particularly well-suited to applications in which users have to attend to several tasks simultaneously, nor for individuals with visual or hearing impairments. Within this context, the introduction of a tactile display, designed to evoke tactile sensation within the skin at the location of the tactile stimulator (e.g., [13]), could present the advantage of freeing visual and auditory channels, by using another unexploited sensory channel to convey information about postural control [48]. Following this train of thought, we developed an original biofeedback system for improving balance control whose underlying principle consists in supplying the user with supplementary sensory information related to foot sole pressure distribution through a tongue placed tactile output device generating electrotactile stimulation of the tongue [39,42,45]. This so-called "Tongue Display Unit" (TDU), initially introduced by Bach-y-Rita et al. [1], comprises a 2D array (1.5 × 1.5 cm) of 36 electrotactile electrodes each of 1.4 mm diameter, arranged in a 6×6 matrix positioned in close contact with the anterior-superior surface of the tongue. In its original version, a flexible cable, passing out of the mouth, connected the matrix to an external electronic device delivering the electrical signals that individually activated the electrodes [39,42,45]. Such architecture, however, ruled out the perspective for an application of this device outside the laboratory framework and its use for long-time periods in a real-life environment. Accordingly, we recently developed, with the help of the companies Coronis-Systems and Guglielmi Technologies Dentaires, a wireless radio-controlled version of this tongue-placed tactile output device including microelectronics, antenna and battery, which can be worn inside the mouth like an orthodontic retainer (Fig. 1). Unfortunately, this wireless TDU allows information related to foot sole pressure distribution to be updated only at a frequency of 3 Hz, in contrast to 50 Hz for the initial wire TDU.

-----------------------------------

Please insert Figure 1 about here

-----------------------------------

The present article describes the architecture and the functioning principle of this new wireless embedded tongue tactile biofeedback system for balance control for fall prevention and presents results of a feasibility study performed on young healthy adults.

## 2. Method

### 2.1. Subjects

Twelve young healthy university students (age: 24.8 ± 4.1 years; body weight: 70.9 ± 12.5 kg; height: 175.7 ± 11.1 cm; mean ± SD) with no history of previous motor problems, neck injury, vertigo, neurological disease, or vestibular impairment voluntarily participated in the experiment. They gave their informed consent to the experimental procedure as required by the Helsinki declaration (1964) and the local Ethics Committee.

### 2.2. Task and procedure

Subjects stood barefoot, feet together, their hands hanging at their sides with their eyes closed. They were asked to sway as little as possible in two no-biofeedback and biofeedback experimental conditions. The no-biofeedback condition served as a control. In the TDU condition, subjects performed a postural task using a plantar pressure-based, tongue-placed tactile biofeedback system. The plantar pressure data acquisition system (Force Sensitive Applications (FSA) Orthotest Mat, Vista Medical Ltd.) was used as the sensory unit. This pressure mat (sensing area: $350 \times 350$ mm = 122 500 mm²), contains a $32 \times 32$ grid of piezo resistive sensors (sensor number: 1024; dimensions: $3.94 \times 3.94$ mm; space between sensors: 2.7 mm; 0.84 sensor/cm²), allowing the magnitude of pressure exerted on each left and right foot sole at each sensor location to be transduced into the calculation of the positions of the resultant centre of foot pressure (CoP) (sampling frequency: 10 Hz). The resultant CoP data were then transmitted to the wireless TDU (Fig. 1) at a frequency of 3 Hz.

-------------------------------------

Please insert Figure 2 about here

-------------------------------------

The underlying principle of this biofeedback system was to supply the user with supplementary information about the position of his/her CoP (white triangles, Fig. 2) relative to a predetermined adjustable "dead zone" (DZ) through the TDU (grey rectangles, Fig. 2) [39,42,45]. In the present experiment, antero-posterior and medio-lateral bounds of the DZ were set as the standard deviation of subject's CoP displacements recorded for 10 s preceding each experimental trial. To avoid an overload of sensory information presented to the user, a simple and intuitive coding scheme for the TDU, consisting of a "threshold-alarm" type of feedback rather that a continuous feedback about the ongoing position of the CoP, was then used:

(1) when the position of the CoP was determined to be within the DZ, no electrical activation was provided in any of the electrodes of the matrix (Fig. 2, central panel);

(2) when the position of the CoP was determined to be outside the DZ – i.e., when it was most needed –, electrical activation of either the anterior, posterior, right or left zone of the matrix ($1 \times 4$ electrodes) (black dots, Fig. 2) (i.e. electrotactile stimulation of front, rear, right of left portion of the tongue) was provided, depending on whether the actual position of the CoP was in an anterior, posterior, right or left position relative to the DZ, respectively (Fig. 2, peripheral panels). For instance, in the case when the CoP was located at the right hand side of the DZ, the activation of four electrodes located in the right portion of the matrix was provided (Fig. 2, right panel). Interestingly, this type of sensory coding scheme for the TDU allows the activation of distinct and exclusive electrodes for a given position of the CoP with respect to the DZ.

Finally, the intensity of the electrical stimulating current was adjusted for each subject, and for each of the front, rear, left, right portions of the tongue, given that the sensitivity to the electrotactile stimulation is known to vary between individuals [6], but also as a function of location on the tongue [40].

Several practice runs were performed prior to the test to ensure that subjects had mastered the relationship between the position of the CoP relative to the DZ and lingual stimulations. Three 30 seconds trials for each experimental condition were performed. The order of presentation of the two experimental conditions was randomized.

## 2.3. Data analysis

Two dependent variables were used to describe the subject's postural behaviour: (1) the surface area (mm²) covered by the trajectory of the CoP with a 85% confidence interval and (2) the range of the CoP displacements (mm). The calculation of the surface area provides a measure of spatial variability of the CoP around the mean position. The range of the CoP displacements indicates the average minimum and maximum excursion of the CoP within the base of supporting any direction. A large value in the range of the CoP displacements indicates that the resultant forces are displaced towards the balance stability boundaries of the subject and could challenge their postural stability.

## 2.4. Statistical analysis

The means of the three trials performed for each of the two experimental conditions were used for statistical analyses. One-way analyses of variance (ANOVAs) 2 Conditions (No-biofeedback vs. Biofeedback) were applied to the data. The level of significance was set at 0.05.

## 3. Results

Figure 3 illustrates representative CoP displacements from a typical participant during standing in the No-biofeedback (A) and Biofeedback (B) conditions.

------------------------------------

Please insert Figure 3 about here

------------------------------------

Analysis of the surface area covered by the trajectory of the CoP shows a main effect of Condition, yielding a smaller value in the Biofeedback than the No-Biofeedback condition $(F(1,11) = 7.34, P < 0.05,$ Fig. 4A).

Results obtained for the range of the CoP displacements are consistent with those obtained for the surface area covered by the trajectory of the CoP. Indeed, the ANOVA confirms the main effect of Condition, yielding a smaller value in the Biofeedback than the No-Biofeedback condition $(F(1,11) = 15.65, P < 0.01,$ Fig. 4B).

On the whole, average reductions of the surface area covered by the trajectory of the CoP and the range of the CoP induced by the use of the biofeedback were 14% and 11%, respectively.

------------------------------------

Please insert Figure 4 about here

------------------------------------

## 4. Discussion

Biofeedback systems for balance control consist in supplying individuals with additional artificial information about body orientation and motion to supplement the natural visual, somatosensory and vestibular sensory cues. Considering that plantar cutaneous information plays an important role in balance control (e.g., [2,14,19,21,23,29,30,31,37]) and that one of the more pervasive effects of aging is loss of cutaneous sensation [15,24,49], we recently developed a biofeedback system whose underlying principle consists in supplying the user with supplementary sensory information related to foot sole pressure distribution through a tongue-placed tactile output device generating electrotactile stimulation of the tongue [39,42,45]. However, to be part of a viable system, a biofeedback system has to be lightweight, portable and capable of several hours of continuous operation, but also aesthetically acceptable. Since the wire TDU did not meet these requirements (a flexible cable connected the matrix of electrodes to an external electronic device), we have developed a wireless embedded tongueplaced tactile output device. It consists of a 2D array electrodes arranged in a $6 \times 6$ matrix glued on to the inferior part of the orthodontic retainer, which also includes microelectronics, antenna and battery. (Fig. 1). The present article describes the architecture and the functioning principle of an original biofeedback system for balance improvement of fall prevention (Fig. 2) and presents results of a feasibility study performed on 12 young healthy adults.

Analyses of the surface area covered by the trajectory of the CoP and the range showed that the postural oscillations were characterized by narrower excursions when biofeedback was in use relative to when it was not (Figs. 3 and 4). These results confirm that electrotactile stimulation of the tongue can be used as part of a biofeedback device designed to improve balance control [39,42,45]. However, it is important to mention that the above-mentioned experiments have used the wire version of the TDU as a tongue-placed tactile output device, which allows the transmission of foot sole pressure distribution to the matrix of electrodes at a 50 Hz frequency. Interestingly, the present findings evidenced the effectiveness of the wireless embedded system, allowing the data to be updated at a 3 Hz frequency only, in improving postural control of young healthy adults. In other words, young healthy adults were able to take advantage of an artificial tongue-placed tactile biofeedback updated every 333 ms to improve their postural control during quiet standing. At this point, it is important to keep in mind that the tongue was chosen as a substrate for an electrotactile stimulation site because of its neurophysiologic characteristics. Indeed, because of its dense mechanoreceptive innervations [35] and large somatosensory cortical representation [26], the tongue can convey higher-resolution information than the skin can [32]. That is certainly one of the reasons why the TDU already has proven its efficiency when used as the sensory output unit for tactile-vision [32], tactile-proprioception [38,40,41] and tactile-vestibular function [36,43,46]. In addition, the high conductivity offered by the saliva insures a highly efficient electrical contact between the electrodes and the tongue surface and therefore does not require high voltage and current [1]. The tongue also is in the protected environment of the mouth and is normally out of sight and out of the way, which could make a tongue-placed tactile display aesthetically acceptable.

Finally, although these feasibility studies have been conducted in young healthy adults, i.e., in individuals with intact sensory, motor, cognitive capacities, we strongly believe that our results could have significant implications in rehabilitative areas, for enhancing/restoring/preserving balance and mobility in individuals with reduced capacities (resulting either from normal aging, trauma or disease) with the aim of ensuring autonomy and safety in occupations of daily living and maximizing the quality of life. Along these lines, the effectiveness of our wireless embedded tongue tactile biofeedback system for balance control is currently being evaluated not only in older healthy adults, but also in individuals with somatosensory loss in the feet from diabetic peripheral neuropathy.

**Acknowledgements**

The authors are indebted to Professor Paul Bach-y-Rita for introducing us to the Tongue Display Unit and for discussions about sensory substitution. Paul has been for us more than a partner or a supervisor: he was a master inspiring numerous new fields of research in many domains of neurosciences, biomedical engineering and physical rehabilitation. This research was supported by the company IDS, Floralis (Université Joseph Fourier, Grenoble) and the Fondation Garches. The company Vista Medical is acknowledged for supplying the FSA pressure mapping system. The authors would like to thank the anonymous reviewers for helpful comments and suggestions. Special thanks also are extended to P. Rainces for various contributions.

## References

[1]   P. Bach-y-Rita, K.A. Kaczmarek, M.E. Tyler, J. Garcia-Lara, Form perception with a 49-point electrotactile stimulus array on the tongue, J. Rehabil. Res. Dev., 35 (1998) 427-430.

[2]   L. Bernard-Demanze, N. Vuillerme, L. Berger, P. Rougier, Magnitude and duration of the effects of plantar sole massages, Int. SportMed J. 7 (2006) 154-169.

[3]   J. Brocklehurst, D. Robertson, J. Groom, Clinical correlates of sway in old age, Age Ageing 11 (1982) 1-9.

[4]   A.J. Campbell, M.J. Borrie, G.F. Spears, S.L. Jackson, J.S. Brown, J.L. Fitzgerald, Circumstances and consequences of falls experienced by a community population 70 years and over during a prospective study, Age Ageing 19 (1990) 136-141.

[5]   L. Chiari, M. Dozza, A. Cappello, F.B. Horak, V. Macellari, D. Giansanti, Audio-biofeedback for balance improvement: an accelerometry-based system, IEEE Trans. Biomed. Eng. 52 (2005) 2108-2111.

[6]   G.K. Essick, A. Chopra, S. Guest, F. McGlone, Lingual tactile acuity, taste perception, and the density and diameter of fungiform papillae in female subjects, Physiol. Behav. 80 (2003) 289-302.

[7]   G.R. Fernie, C.I. Gryfe, P.J. Holliday, A. Llewellyn, The relationship of postural sway in standing to the incidence of falls in geriatric subjects, Age Ageing 1 (1982) 11-16.

[8]   A. Gabell, M.A. Simons, U.S. Nayak, Falls in the healthy elderly: predisposing causes, Ergonomics 28 (1985) 965-975.

[9]   L. Gillespie, W. Gillespie, M. Robertson, S. Lamb, R. Cumming, B. Rowe, Interventions for preventing falls in elderly people, Cochrane Database Syst. Rev. 3 (2001). CD000340.

[10] F. Horak, C. Shupert, A. Mirka, Components of postural dyscontrol in the elderly: a review, Neurobiol. Aging, 10 (1989) 727-745.

[11] J. Hegeman, F. Honneger, M. Kupper, J.H. Allum, The balance control of bilateral peripheral vestibular loss subjects and its improvement with auditory prosthetic feedback, J. Vest. Res. 15 (2005) 109-117.

[12] P.O. Jantti, I. Pyykko, P. Laippala, Prognosis of falls among elderly nursing home residents, Aging (Milano), 7 (1995) 23-27.

[13] K.A. Kaczmareck, J.G. Webster, P. Bach-y-Rita, W.J. Tompkins, Electrotactile and vibrotactile displays for sensory substitution systems, IEEE Trans. Rehabil. Eng. 38 (1991) 1-16.

[14] A. Kavounoudias, R. Roll, J.P. Roll, The plantar sole is a "dynamometric map" for human balance control, NeuroReport 9 (1998) 3247-3252.

[15] D.R. Kenshalo, Somesthetic sensitivity in young and elderly humans, J. Gerontol., 41 (1986) 632-642.

[16] K. Lindqvist, T. Timpka, L. Schlep, Evaluation of an inter-organizational prevention program against injuries among the elderly, Public Health, 115 (2001) 308-316.

[17] S.R. Lord, R.D. Clark, I.W. Webster, Physiological factors associated with falls in an elderly population J. Am. Geriatr. Soc. 39 (1991) 1194-1200.

[18] S.R. Lord, H.B. Menz, A. Tiedemann, A physiological profile approach to falls risk assessment and prevention, Phys. Ther. 83 (2003) 237-252.

[19] B.E. Maki, S.D. Perry, R.G. Norrie, W.E. McIlroy, Effect of facilitation of sensation from plantar foot-surface boundaries on postural stabilization in young and older adults, J. Gerontol. A Biol. Sci. Med. Sci. 54 (1999) M281-M287.

[20] J. Massion, Postural control system, Curr. Opin. Neurobiol. 4 (1994) 877-887.

[21] C. Maurer, T. Mergner, B. Bolha, F. Hlavacka, Human balance control during cutaneous stimulation of the plantar soles, Neurosci. Lett. 302 (2001) 45-48.

[22] H.B. Menz, M.E .Morris, S.R. Lord, Foot and ankle risk factors for falls in older people: a prospective study, J. Gerontol. A Biol. Sci. Med. Sci., 61 (2006) 866-870.

[23] P.F. Meyer, L.I. Oddsson, C.J. De Luca, The role of plantar cutaneous sensation in unperturbed stance, Exp. Brain Res., 156 (2004) 505-512.

[24] S.D. Perry, Evaluation of age-related plantar-surface insensitivity and onset age of advanced insensitivity in older adults using vibratory and touch sensation tests, Neurosci. Lett. 392 (2006) 62-67.

[25] H. Petersen, M. Magnusson, R. Johansson, P.A. Fransson, Auditory feedback regulation of perturbed stance in stroke patients, Scand. J. Rehabil. Med. 28 (1996) 217-223.

[26] C. Picard, A. Olivier, Sensory cortical tongue representation in man, J. Neurosurg. 59 (1983) 781-789.

[27] M. Piirtola, P. Era, Force platform measurements as predictors of falls among older people - a review, Gerontology. 52 (2006) 1-16. Review.

[28] N. Pinsault, N. Vuillerme, The effects of scale display of visual feedback on postural control during quiet standing in healthy elderly subjects, Arch. Phys. Med. Rehabil. (2008) (in press).

[29] A. Priplata, J. Niemi, M. Salen, J. Harry, L.A. Lipsitz, J.J. Collins, Noise-enhanced human balance control, Phys. Rev. Lett. 89 (2002) 238101.

[30] A. Priplata, J. Niemi, M. Salen, J. Harry, L.A. Lipsitz, J.J. Collins, Vibrating insoles and balance control in elderly people, Lancet 362 (2003) 1123-1124.

[31] M.S. Redfern, P.L. Moore, C.M. Yarsky, The influence of flooring on standing balance 13 among older persons, Hum. Factors 39 (1997) 445-455.

[32] E. Sampaio, S. Maris, P. Bach-y-Rita, Brain plasticity: Visual acuity of blind persons via the tongue, Brain Res. 908 (2001) 204-207.

[33] A. Shumway-Cook, D. Anson, S. Haller, Postural sway biofeedback: its effect on reestablishing stance stability in hemiplegic patients, Arch. Phys. Med. Rehabil. 69 (1988) 395-400.

[34] M.E. Tinetti, M. Speechley, S.F. Ginter, Risk factors for falls among elderly persons living in community, N. Engl. J. Med. 319 (1988) 1011-1017.

[35] M. Trulsson, G.K. Essick, Low-threshold mechanoreceptive afferents in the human lingual nerve, J. Neurophysiol. 77 (1997) 737-748.

[36] M. Tyler, Y. Danilov, P. Bach-y-Rita, P., Closing an open-loop control system: vestibular substitution through the tongue, J. Integr. Neurosci. 2 (2003) 159-164.

[37] J. Vaillant, N. Vuillerme, A. Janvy, F. Louis, R. Braujou, R. Juvin, V. Nougier, Effect of manipulation of the feet and ankles on postural control in elderly adults, Brain Res. Bull. 75 (2008) 18-22.

[38] N. Vuillerme, M. Boisgontier, O. Chenu, J. Demongeot, Y. Payan, Tongue-placed tactile biofeedback suppresses the deleterious effects of muscle fatigue on joint position sense at the ankle, Exp. Brain Res. 183 (2007a) 235-240.

[39] N. Vuillerme, O. Chenu, J. Demongeot, Y. Payan, Controlling posture using a plantar pressure-based, tongue-placed tactile biofeedback system, Exp. Brain Res. 179 (2007b) 409-414.

[40] N. Vuillerme, O. Chenu, J. Demongeot, Y. Payan, Improving human ankle joint position sense using an artificial tongue-placed tactile biofeedback, Neurosci. Lett. 405 (2006a) 19-23.

[41] N. Vuillerme, O. Chenu, A. Fleury, J. Demongeot, Y. Payan, Optimizing the use of an artificial tongue-placed tactile biofeedback for improving ankle joint position sense in

humans, in: 28th Annual International Conference of the IEEE Engineering In Medicine and Biology Society (EMBS), New York, USA, (2006b). 6029-6032.

[42]  N. Vuillerme, O. Chenu, N. Pinsault, M. Boisgontier, J. Demongeot, Y. Payan, Inter-individual variability in sensory weighting of a plantar pressure-based, tongue-placed tactile biofeedback for controlling posture, Neurosci. Lett. 421 (2007c) 173-177.

[43]  N. Vuillerme, R. Cuisinier, Head position-based electrotactile tongue biofeedback affects postural responses to Achilles tendon vibration in humans. Exp. Brain Res. 186 (2008) 503-508.

[44]  N. Vuillerme, N. Pinsault, Re-weighting of somatosensory inputs from the foot and the ankle for controlling posture during quiet standing following trunk extensor muscles fatigue, Exp. Brain Res. 183 (2007) 323-327.

[45]  N. Vuillerme, N. Pinsault, O. Chenu, M. Boisgontier, J. Demongeot, Y. Payan, How a plantar pressure-based, tongue-placed tactile biofeedback modifies postural control mechanisms during quiet standing, Exp. Brain Res. 181 (2007d) 547-554.

[46]  N. Vuillerme, N. Pinsault, O. Chenu, J. Demongeot, Y. Payan, Y. Danilov, Sensory supplementation system based on electrotactile tongue biofeedback of head position for balance control. Neurosci. Lett. 431 (2008) 206-210.

[47]  N. Vuillerme, N. Pinsault, J. Vaillant, Postural control during quiet standing following cervical muscular fatigue: effects of changes in sensory inputs, Neurosci. Lett. 378 (2005) 135-136.

[48]  C. Wall, M.S. Weinberg, P.B. Schmidt, D.E. Krebs, Balance prosthesis based on micromechanical sensors using vibrotactile feedback of tilt, IEEE Trans. Biomed. Eng. 48 (2001) 1153-1161.

[49]  C. Wells, L.M. Ward, R. Chua, J.T. Inglis, Regional variation and changes with ageing in vibrotactile sensitivity in the human footsole, J. Gerontol. A Biol. Sci. Med. Sci. 58 (2003) 680-686.

[50]  G. Wu, Real-time feedback of body center of gravity for postural training of elderly patients with peripheral neuropathy, IEEE Trans. Rehabil. Eng. 5 (1997) 399-402.

**Figure captions**

**Figure 1.** Photograph of the wireless radio-controlled tongue-placed tactile output device developed in the TIMC Laboratory. It consists in a 2D electrodes array arranged in a $6 \times 6$ matrix glued onto the inferior part of the orthodontic retainer which also includes microelectronics, antenna and battery.

**Figure 2.** Sensory coding schemes for the Tongue Display Unit (TDU) as a function of the position of the centre of foot pressure (CoP) relative to a predetermined dead zone (DZ).

White triangles, grey rectangles and black dots represent the positions of the CoP, the predetermined dead zones and activated electrodes, respectively.

There were 5 possible stimulation patterns of the TDU.

On the one hand, no electrodes were activated when the CoP position was determined to be within the DZ (*central panel*).

On the other hand, 4 electrodes located in the front, rear, left and right zones of the $6 \times 6$ matrix were activated when the CoP positions were determined to be outside the DZ, located towards the front, rear, left and right of the DZ, respectively (*peripheral panels*). These 4 stimulation patterns correspond to the stimulations of the front, rear, left and right portions of the tongue dorsum, respectively.

**Figure 3.** Representative displacements of the centre of foot pressure (CoP) from a typical subject recorded in the No-biofeedback (left panel) and Biofeedback (right panel) conditions.

**Figure 4.** Mean and standard deviation of the standard deviation of the surface area covered by the CoP (A) and the range of the CoP displacements obtained in the two No-biofeedback

and Biofeedback conditions. The two conditions of No-biofeedback and Biofeedback are presented with different symbols: No-biofeedback (*white bars*) and Biofeedback (*black bars*). The significant P-values for comparison between No-biofeedback and Biofeedback conditions also are reported (*: $P < 0.05$, **: $P < 0.01$).

**Figure 1**

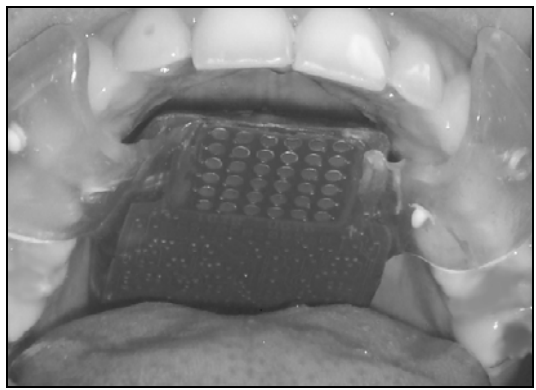

**Figure 2**

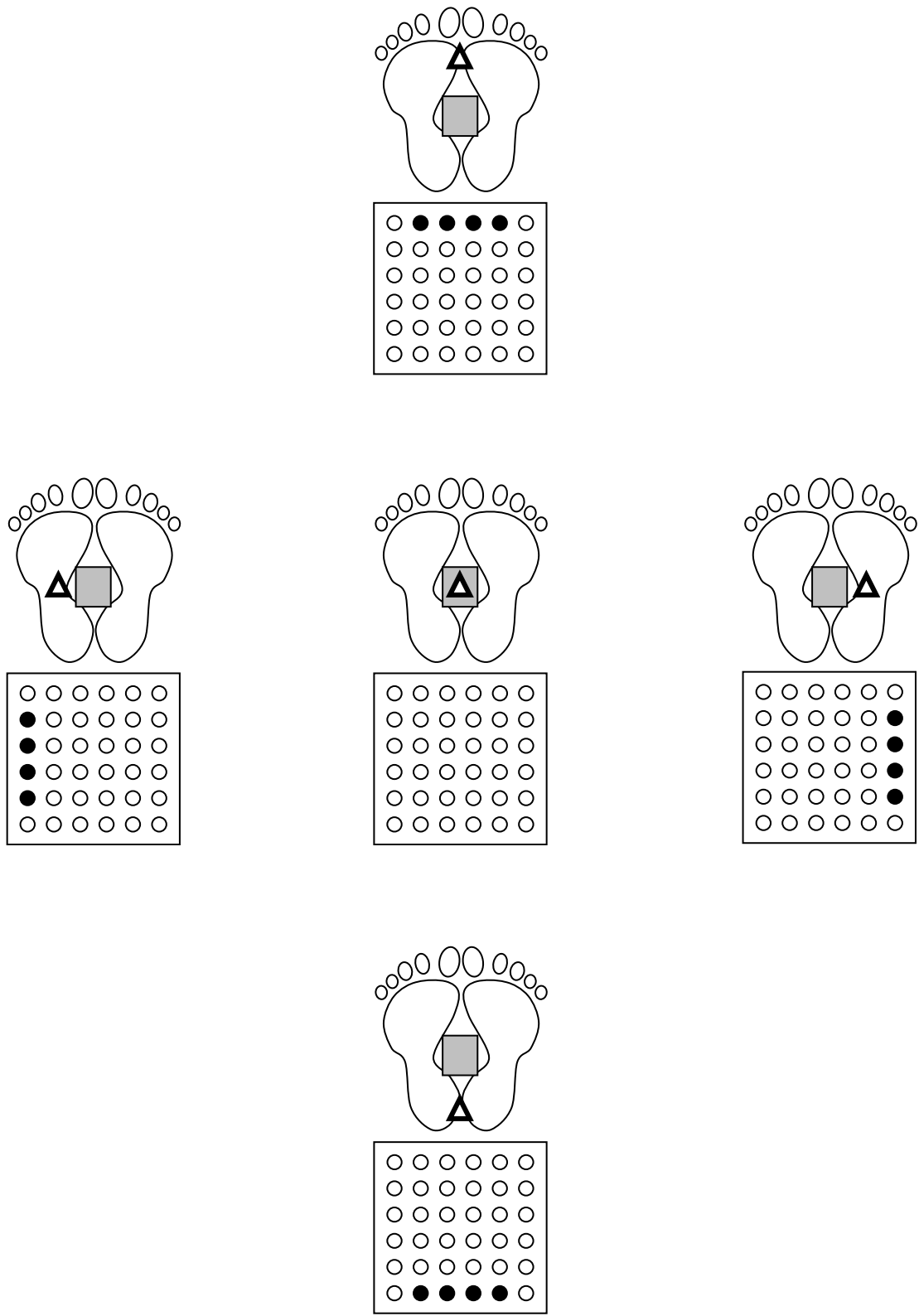

**Figure 3**

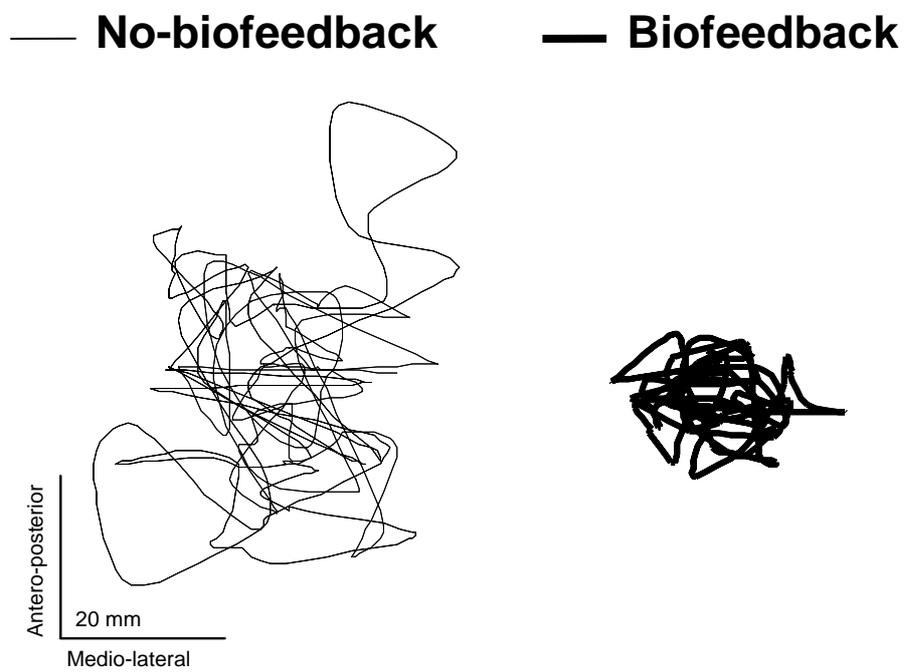

**Figure 4**

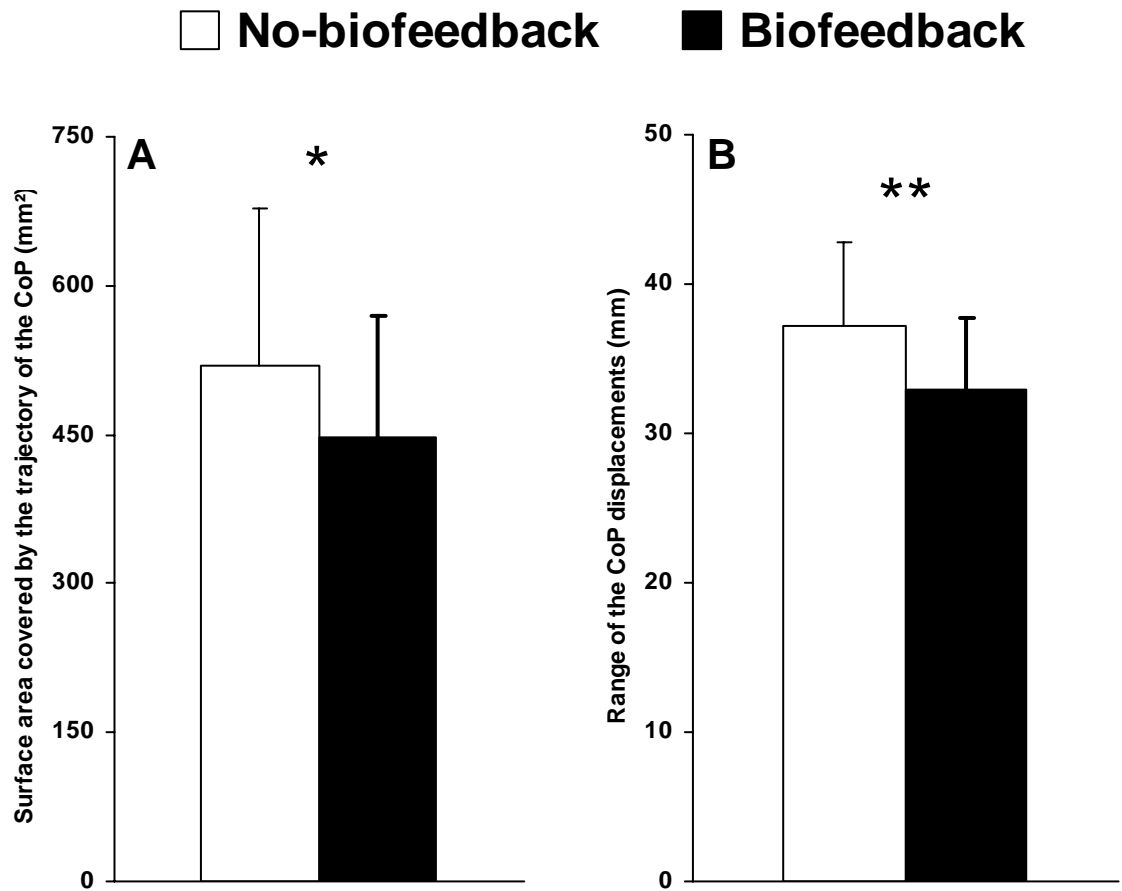